\documentstyle[preprint,aps,epsfig]{revtex}

 \newcommand{\mg}{m_{\tilde{g}}}

\def\bea{\begin{eqnarray}}
\def\eea{\end{eqnarray}}

\def\ra{\rightarrow}

\def\chargino{{\tilde{\chi}}^{\pm}}
\def\neutralino{{\tilde{\chi}}^{0}}
\def\gaugino{{\tilde{\chi}}}
\def\gluino{\tilde{g}}

\def\w3ino{\tilde{W_3}}
\def\h1ino{\tilde{H_1}}
\def\hino2{\tilde{H_2}}

\def\alphas{\alpha_{S}}
\def\alphash{\hat{\alpha}_{S}}
\def\MS{$\overline{\rm MS}$\,\,}

\def\u6d{u_{6 \Delta}}
\def\ud7{u_{7 \Delta}}
\def\s4d{s_{4 \Delta}}
\def\sd3{s_{3 \Delta}}

\newcommand{\msqu}[1]{m_{\tilde{q}_{#1}}}

\begin{document}
\draft

\preprint{
\vbox{
\halign{&##\hfil\cr
	& ANL-HEP-PR-99-03 \cr
	& MSUHEP-90215 \cr}}
}
\title{Associated Production of Gauginos and Gluinos at \\
       Hadron Colliders in Next-to-Leading Order SUSY-QCD}
\author{Edmond L. Berger$^a$, Michael Klasen$^a$, and Tim Tait$^{a,b}$}
\address{$^a$Argonne National Laboratory,
             Argonne, Illinois 60439 \\
         $^b$Michigan State University,
             East Lansing, Michigan 48824}
\maketitle

\begin{abstract} 
We report a next-to-leading order (NLO) calculation of the production of
gaugino-like charginos ($\chargino$) and neutralinos ($\neutralino$) in
association with gluinos ($\gluino$) at hadron colliders, including the
strong corrections from colored particles and sparticles. We predict
inclusive cross sections at the Fermilab Tevatron and CERN LHC. The NLO cross 
sections are more stable against variations in the hard-scattering scale 
parameter and are greater than the LO values.

\end{abstract} 
\vspace{0.2in}


Supersymmetry predicts the existence of supersymmetric partners for each of the 
particles of the standard model.  The search for these sparticles is a 
principal motivation of the forthcoming Run II of the Fermilab Tevatron 
collider and of the CERN Large Hadron Collider (LHC) program.  A potentially 
important, but heretofore largely overlooked, discovery channel is the 
associated production of a spin-1/2 gaugino ($\gaugino$) with a spin-1/2 gluino 
($\gluino$) or with a spin-0 squark ($\tilde{q}$).  Color-neutral gauginos 
couple with electroweak strength, whereas the colored squarks and gluinos 
couple strongly.  Associated production is therefore a semi-weak process in 
that it involves one somewhat smaller coupling constant than the pair 
production of colored sparticles.  However, in popular models of SUSY 
breaking \cite{sugra,gaugem}, the mass spectrum favors much lighter masses for 
the low-lying neutralinos and charginos than for the squarks and gluinos.  
This mass hierarchy means that the phase space for production of neutralinos 
and charginos, the corresponding partonic luminosities, and the 
production cross sections will be greater than those for gluinos and squarks.  
These advantages are potentially decisive at a collider with limited energy, 
such as the Tevatron. Furthermore, associated production has a clean 
experimental signature.  For example, the lowest lying neutralino is the 
(stable) lightest supersymmetric particle (LSP) in supergravity (SUGRA) 
models \cite{sugra}, manifest as missing energy in the events, and it is the 
second lightest in gauge-mediated models \cite{gaugem}.  In models with a very 
light gluino \cite{farraraby}, there could be large rates for 
$\gluino \gaugino$ production, with simple signatures, whereas 
$\gluino \gluino$ production suffers from large hadronic jet backgrounds.   

Experimental investigations are facilitated by firm theoretical understanding 
of the expected sizes of the cross sections for production of the
superparticles.  In the case of hadron-hadron colliders,
the large strong coupling strength ($\alphas$) results in potentially
large contributions to cross sections from terms beyond leading order
(LO) in a perturbative quantum chromodynamics (QCD) evaluation of the cross 
section.  For accurate theoretical estimates, it is necessary to extend the 
calculations to next-to-leading order (NLO) or beyond.  NLO contributions 
generally reduce and stabilize dependence on undetermined parameters such 
as the renormalization and factorization scales.  To date, associated 
production has been calculated only in LO \cite{lo}, but NLO results exist for 
hadroproduction of gluinos and squarks \cite{squarkgluino}, top 
squarks \cite{stop}, sleptons \cite{slepton,gaugino}, and 
gauginos \cite{gaugino}.  Studies have begun to incorporate these NLO results 
into Monte Carlo simulations \cite{scales}.

In this Letter we present the first NLO (in SUSY-QCD) calculation of 
hadroproduction of a $\gluino$ in association with a $\gaugino$, including 
contributions from virtual loops of colored sparticles and particles and 
three-particle final states involving the emission of light real particles.  
We extract the ultraviolet, infrared, and collinear divergences by use of 
dimensional regularization and employ standard \MS renormalization and mass 
factorization procedures.  In the course of computing the virtual 
contributions, we encountered new divergent four-point functions.  The 
contributions from real emission of light particles are treated with a phase 
space slicing method.  We provide predictions for inclusive cross sections at 
Tevatron and LHC energies. Our reason to focus on the $\gluino + \gaugino$ 
final state, rather than on the associated production $\tilde{q} + \gaugino$, 
is that at the energy of the Tevatron the LO cross sections for 
$\gluino + \gaugino$ are 3 to 6 times greater than those for 
$\tilde{q} + \gaugino$ ($\gaugino = \gaugino^0_{1,2}, \gaugino^{\pm}_1$) when  
$m_{\gluino} = m_{\tilde{q}} = 300$ GeV, and 6 to 15 times greater when 
$m_{\gluino} = m_{\tilde{q}} = 600$ GeV.  In obtaining the $\tilde{q}$ cross 
sections, we sum over five flavors of squarks and antisquarks.  

In LO of SUSY-QCD, the associated production of a gluino and a gaugino
proceeds through the subprocess $q\bar{q}\ra\gluino\gaugino$ with a $t$-channel
or a $u$-channel squark exchange. We assume that there is no mixing between
squarks of different generations and that the squark mass eigenstates are
aligned with the squark chirality states, equivalent to the assumption that 
the two squarks of a given flavor are degenerate in mass. We ignore the 
$n_f = 5$ light quark masses in all of the kinematics and couplings.  Under 
these assumptions, the massless incoming quarks and antiquarks have a 
particular helicity, and thus the Feynman diagrams in which a
right-handed squark is exchanged cannot interfere with those mediated
by a left-handed squark. In evaluating the Feynman diagrams involving
Majorana and explicitly charge-conjugated fermions, we follow
the approach of Ref.\cite{denner}. In the case of charged gauginos,
only the left-handed squarks participate, whereas neutral gauginos receive
contributions from both left- and right-handed squarks.

The LO matrix element summed (averaged) over the colors and 
helicities of the outgoing (incoming) particles has the analytic form \cite{lo}
\bea
 \overline{|{\cal M}^B|}^{2} = \frac{ 8 \pi\, \alphash}{9} &
 \Biggl[ & \frac{\hat{X}_t\, t_{\gluino}\, t_{\gaugino}}{(t - \msqu{t}^2)^2}
 - \frac{2 \, \hat{X}_{tu}\, s\, m_{\gluino}\, m_{\gaugino}}{(t - \msqu{t}^2)
 (u - \msqu{u}^2)} \nonumber \\
 & + & \frac{\hat{X}_u\, u_{\gluino}\, u_{\gaugino}}{(u - \msqu{u}^2)^2}
 \Biggr] . \label{loeq} 
\eea
Here, $\msqu{t,u}$ is the mass of the squark exchanged in the $t$- and
$u$-channel; $\alphash =\hat{g}^2_S/4\pi$ is the Yukawa coupling between
quarks, squarks, and gluinos.  At leading order it is equal to the gauge
coupling constant $\alphas$.  In Eq.~(\ref{loeq}) $\hat{X}_{t,tu,u}$ stands 
for the Yukawa couplings of quarks, squarks, and gauginos.   
Variables $s$, $t$, and $u$ are the usual Mandelstam invariants at the 
partonic level; $t_{\gluino,\gaugino}=t-m_{\gluino,\gaugino}^2$ and 
$u_{\gluino,\gaugino}=u-m_{\gluino,\gaugino}^2$. 

At NLO the cross section receives contributions from virtual loop diagrams
and from real parton emission diagrams. The virtual contributions arise from
the interference of the Born amplitudes with the related one-loop amplitudes
containing self-energy corrections, vertex corrections, and box diagrams.
We include the full supersymmetric spectrum of strongly interacting particles
in the virtual loops, i.e.\ squarks and gluinos as well as quarks and gluons.

Since the virtual loop contributions are ultraviolet and infrared divergent,
we regularize the cross section by computing the phase space and matrix
elements in $n=4-2\epsilon$ dimensions.
We calculate the traces of Dirac matrices using the ``naive'' $\gamma_5$
scheme in which $\gamma_5$ anticommutes with all other $\gamma_{\mu}$
matrices. This choice is justified for anomaly-free one-loop amplitudes.
The $\gamma_5$ matrix enters the calculation through both the
quark-squark-gluino and quark-squark-gaugino Yukawa couplings.
We simplify the integration over the internal loop momenta by
reducing all tensorial integration kernels to expressions that are only
scalar in the loop momentum \cite{pasvelt}.
The resulting one-, two-, three-, and some of the four-point functions were 
computed in the context of other physical processes \cite{squarkgluino}.
However, we compute two previously unknown divergent four-point functions; 
these new functions arise because the final state gluino and gaugino have 
different masses in general.  We evaluate the scalar four-point functions 
by calculating the absorptive parts with Cutkosky cutting rules and the real 
parts with dispersion techniques.

The ultraviolet (UV) divergences are manifest in the one- and two-point 
functions as poles in $1/\epsilon$.  We remove them by renormalizing the 
coupling constants in the $\overline{\rm MS}$ scheme at the renormalization 
scale $Q$ and the masses of the heavy particles (squarks and gluinos) in the 
on-shell scheme.  The self-energies for external particles are multiplied by 
a factor of $1/2$ for proper wave function renormalization.  A difficulty 
arises from the fact that spin-1 gluons have $n-2$ possible polarizations, 
whereas spin-1/2 gluinos have 2, leading to broken supersymmetry in the 
$\overline{\rm MS}$ scheme.  The simplest procedure to restore supersymmetry 
is with finite shifts in the quark-squark-gluino and quark-squark-gaugino 
Yukawa couplings~\cite{mismatch,squarkgluino}.

In addition to the ultraviolet singularities, the virtual corrections have
collinear and infrared singularities that show up as $1/\epsilon$ or
$1/\epsilon^2$ poles in the derivatives of the two-point function and in the 
three- and four-point functions.
These infrared singularities appear as factors times parts of the Born matrix
elements. They can be separated into $C_F$ and $N_C$ color classes, depending
on the color flow and the abelian or non-abelian nature of the correction
vertices. They are cancelled eventually by corresponding soft and collinear
singularities from the real three parton final state corrections.

The real corrections to the production of gluinos and gauginos arise from 
three parton final-state subprocesses in which additional
gluons and massless quarks and antiquarks are emitted: 
$q\bar{q}\ra\gluino\gaugino g$, $qg\ra\gluino\gaugino q$, and
$\bar{q}g\ra\gluino\gaugino\bar{q}$. 

The $n$-dimensional phase space for $2\ra 3$ scattering may be 
factored into the phase space for $2\ra 2$ scattering and the phase space 
for the subsequent decay of one of the two final state particles with squared 
invariant mass $s_4=(p_1+p_3)^2-m_1^2$ into two particles with momenta
$p_1$ and $p_3$, parametrized in the rest frame of particles
1 and 3 \cite{hquark}.
We follow the procedure in Ref.\cite{hquark} to reduce all of the
angular integrals.
The angular integrations involving negative powers
of $t'=(p_b-p_3)^2$ and $u'=(p_a-p_3)^2$, where $p_a$ and $p_b$ are the
four-momenta of the incoming partons, produce poles in $1/\epsilon$ which
correspond to the collinear singularities in which particle 3 is collinear
with particle $a$ or $b$.  
Because these singularities have a universal structure,
they may be removed from the cross section and absorbed into the parton
distribution functions according to the usual mass factorization procedure
\cite{facto}.

In addition to the collinear singularities described above, the corrections
involving real gluon emission also have infrared (IR) singularities arising
when the energy of the emitted gluon approaches zero.  These singularities
appear as poles in $s_4$ in the cross section and must also be extracted so
that they can be combined with corresponding terms in the virtual corrections
and shown to cancel. In order to make this cancellation conveniently, we
slice the gluon emission phase space into hard and soft pieces,
\begin{equation}
  \label{softhard}
  \frac{d^2\hat{\sigma}_{ij}^R}{dt_\gaugino du_\gaugino} =
  \int_0^{\Delta} ds_4 \,\frac{d^3\sigma^S}{dt_\gaugino\, du_\gaugino\, ds_4} +
  \int_\Delta^{{s_4}^{max}} ds_4 \,\frac{d^3\sigma^H}{dt_\gaugino\,
  du_\gaugino \, ds_4},
\end{equation}
where $\Delta$ is an arbitrary cut-off between soft and hard
gluon radiation. When the cut-off is much smaller than the other invariants,
the $s_4$ integration for the soft term becomes simple and can be
evaluated analytically, leading to explicit logarithms $\log\Delta/m^2,
\log^2\Delta/m^2$; $m = (m_{\gluino}+m_{\gaugino})/2$. The hard term is free 
from infrared and, after mass
factorization, also collinear singularities and can be evaluated numerically
in four dimensions. This procedure leads to an implicit logarithmic dependence 
of the hard term on the cut-off $\Delta$ which cancels the explicit logarithmic
dependence in the soft term.

To obtain numerical results for the cross sections, we work within a 
particular SUGRA scheme, but the cross sections depend principally on the 
masses of the $\gaugino$ and $\gluino$ and are otherwise fairly independent of 
the details of the SUSY breaking.  
The physical gluino and gaugino masses as well as the gaugino mixing matrices
are calculated from a default minimal SUGRA scenario \cite{sugra}. We choose
the common scalar and fermion masses at the GUT scale to be $m_0 = 100$ GeV
and $m_{1/2} = 150$ GeV. The trilinear coupling $A_0 = 300$ GeV, and the
ratio of the Higgs vacuum expectation values $\tan\beta = 4$. The absolute value
of the Higgs mass parameter $\mu$ is fixed by electroweak symmetry breaking,
and we choose $\mu > 0$. For this scenario, we find the neutralino masses
$m_{\neutralino_{1-4}}$ to be 55, 104, 283, and 309 GeV with
$m_{\neutralino_3} < 0$ inside a polarization sum. The chargino masses
$m_{\chargino_{1,2}}$ are 102 and 308 GeV and therefore almost degenerate
with the masses of the $m_{\neutralino_2}$ and $m_{\neutralino_4}$, 
respectively.  

The total hadronic cross section is obtained from the
partonic cross section through the convolution 
\bea
 \sigma^{h_1h_2}(S,Q^2) &=& \sum_{i,j=g,q,\overline{q}}
 \int_{\tau}^1{\rm d}x_1\int_{\tau/x_1}^1{\rm d}x_2 \nonumber\\
 && f_i^{h_1}(x_1,Q^2) f_j^{h_2}(x_2,Q^2)
 \hat{\sigma}_{ij}(x_1x_2S,Q^2), 
\label{e4}
\eea
where $\tau = \frac{4m^2}{S}$, 
and $S$ is the square of the hadronic center-of-mass energy ($\sqrt S = $  
2 TeV for Run II at the Fermilab $p\bar{p}$ collider Tevatron and 14 TeV at 
the CERN LHC $pp$ collider). For the NLO predictions, we employ the CTEQ4M 
parametrization \cite{cteq4} for the parton densities $f(x,Q^2)$ in
the proton or antiproton and a two-loop approximation for the strong
coupling constant $\alphas$ with $\Lambda^{(5)}=202$ MeV. In LO we use the
LO parton densities CTEQ4L and the one-loop approximation for $\alphas$
with $\Lambda^{(5)}=181$ MeV.  

\begin{figure}
 \begin{center}
  \vspace*{-10mm}
  \epsfig{file=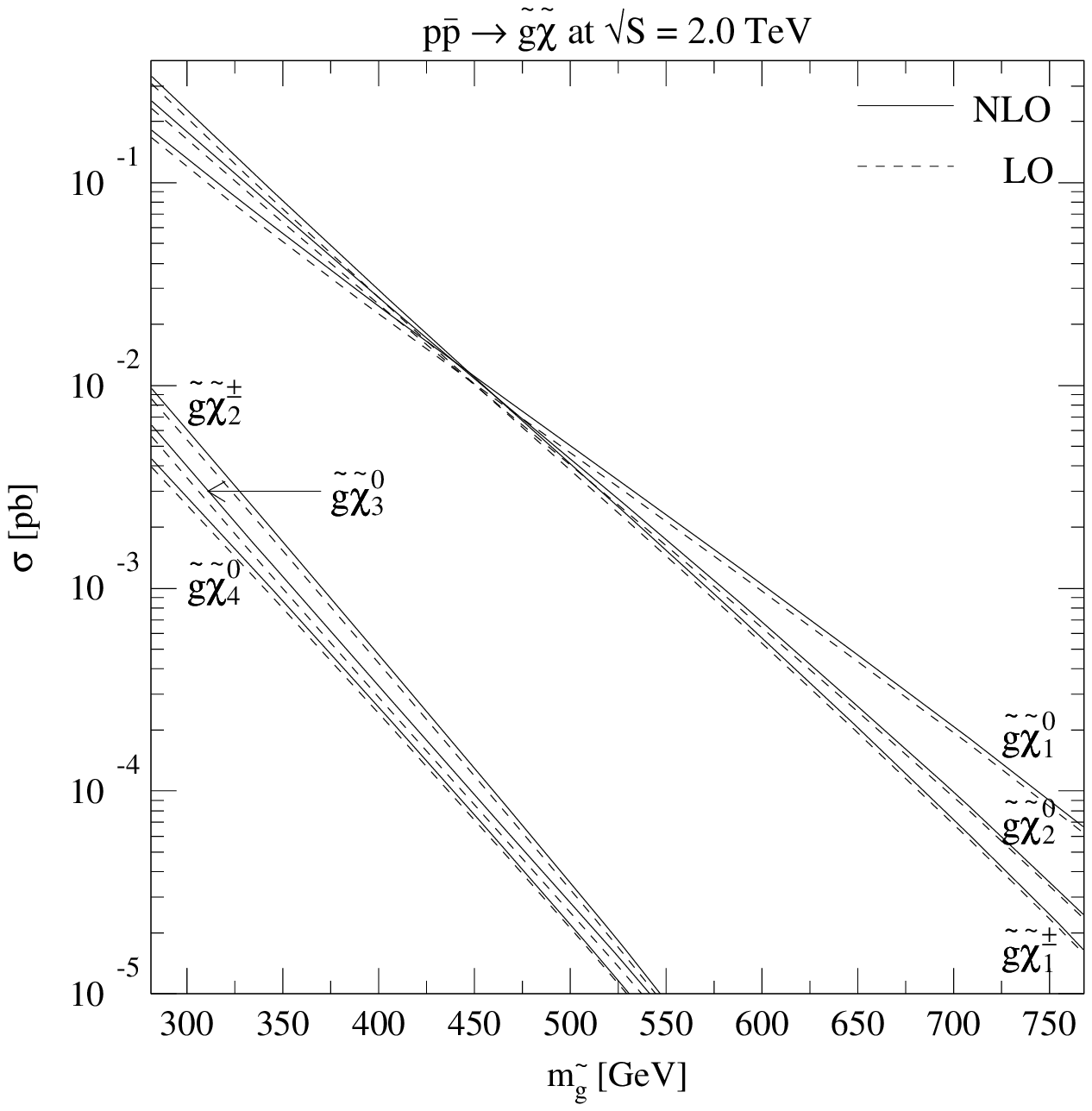,width=10cm}
  \vspace*{-4mm}
  \epsfig{file=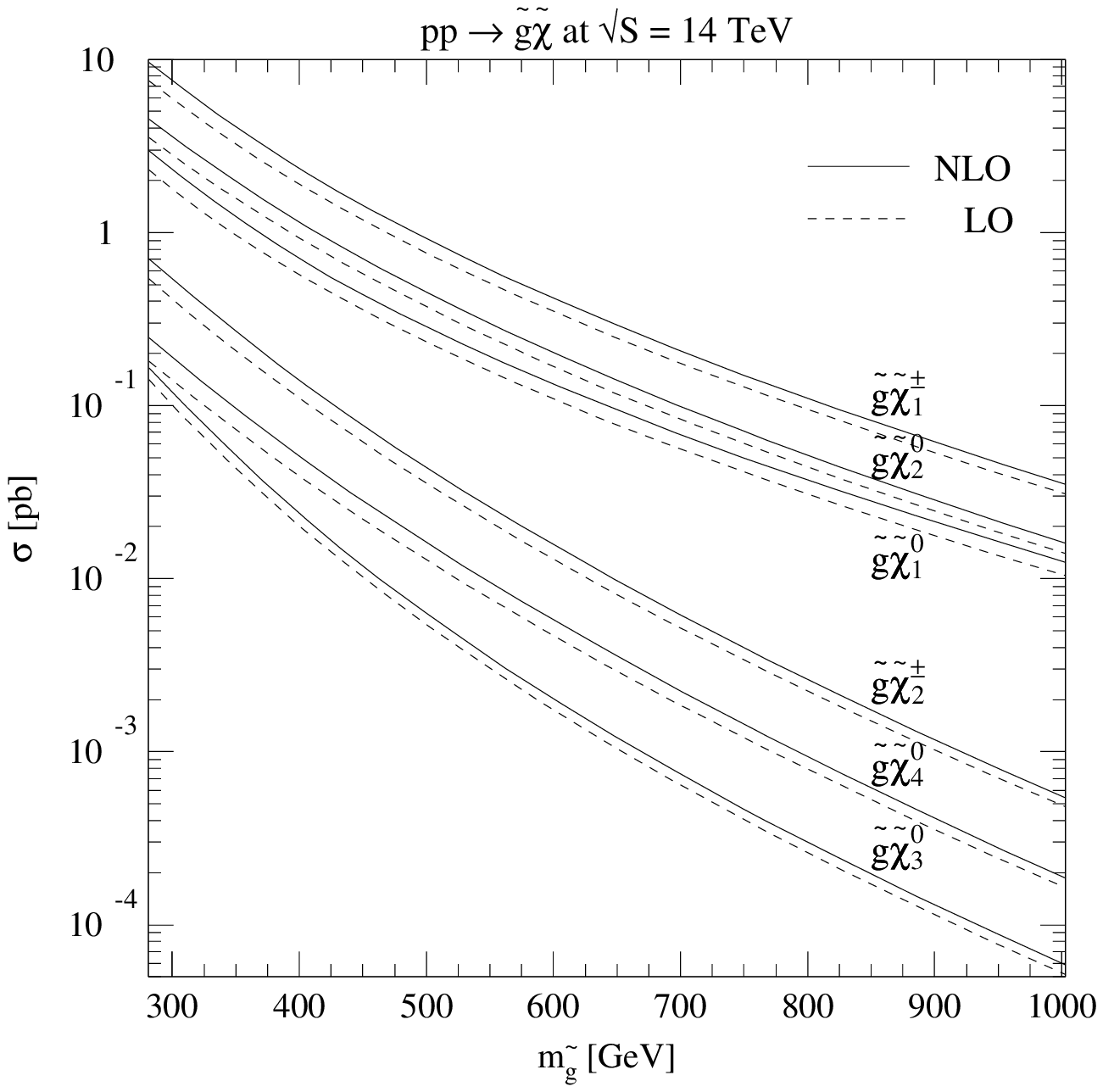,width=10cm}
 \end{center}
 \caption{Total hadronic cross sections for the associated production
  of gluinos and gauginos at Run II of the Tevatron (top) and the LHC
  (bottom).  NLO results are shown as solid curves, and LO results as 
  dashed curves.  We vary the SUGRA scenario as a function of $m_{1/2} \in
  [100;400]$ GeV and provide the cross sections as a function of the
  physical gluino mass $\mg$. The chargino cross sections are summed
  over positive and negative chargino charges.}
\label{xsec}
\end{figure}

In Fig. \ref{xsec} we present predictions for total hadronic cross sections 
at the Tevatron (top) and the LHC (bottom), as a function of the physical 
gluino mass.  To obtain these results, we use the average produced mass as the 
hard scale $Q$ in Eq.(\ref{e4}), $Q = (m_{\tilde g} + m_{\tilde \chi})/2$.  We 
vary the SUGRA parameter $m_{1/2}$ between 100 and 400 GeV 
and keep the other SUGRA parameters fixed to the values listed above.  As 
gluino mass increases over the range shown in the figure, the corresponding 
gaugino mass ranges are
31 to 163 GeV for $\neutralino_1$, 62 to 317 GeV for $\neutralino_2$ and
$\chargino_1$, 211 to 666 GeV for $\neutralino_3$, and 240 to 679 GeV for
$\neutralino_4$ and $\chargino_2$. The chargino cross sections are summed over
positive and negative charges. 

We observe that the cross sections for $\neutralino_2$ and $\chargino_1$ and 
those for $\neutralino_4$ and $\chargino_2$ are very similar in magnitude at 
the Tevatron, as are their respective masses. One might expect the largest 
cross section
for the lightest gaugino $\neutralino_1$. However, its coupling is dominantly
photino-like and smaller than the zino-like coupling of $\neutralino_2$ which 
therefore has a larger cross section at small $\mg$ despite its larger
mass. The heavier gauginos $\neutralino_{3,4}$ and $\chargino_2$ are dominantly
higgsino-like and their cross sections are suppressed by over an order of 
magnitude with respect to those of the lighter gauginos.

At the Tevatron, the NLO contributions increase the cross sections by 5 to 
15\% at the hard scattering scale $Q = (m_{\tilde g} + m_{\tilde \chi})/2$, 
depending on the channel considered and the values of the masses.  At 
the LHC, the increases are in the range of 15 to 35\%.   The purely NLO $q g$ 
incident channel contributes significantly at the LHC, in addition to 
the $q \bar{q}$ channel, particularly for the lighter gauginos, whereas 
the $q g$ channel plays little role at the Tevatron.  In the event sparticles 
are not observed, the predicted increases translate into more restrictive 
experimental mass limits.

The enhancements of the cross sections are modest and, as such, underscore the 
validity of perturbative predictions for the processes considered.  A further 
important benefit of the NLO computation is the considerable reduction in 
theoretical uncertainty associated with variation of the renormalization
and factorization scale $Q$.  For the processes studied here, this dependence 
is typically $\pm 10$\% at the Tevatron when $Q$ is varied over the interval 
$Q/m$ from 0.5 to 2, compared to $\pm 25$\% in leading order.  At the LHC, 
the dependences are $\pm 9$\% at NLO and $\pm 12$\% at LO.  

In this Letter we limit ourselves to total cross sections. Differential
distributions in the transverse momentum $p_T$ and the rapidity $\eta$ of
the produced sparticles will be published elsewhere \cite{longpaper}, along 
with figures of scaling functions, renormalization/factorization scale 
dependence, K-factors, and several appendices containing a detailed exposition 
of the calculation. 

To summarize, we provide NLO predictions of the cross sections for 
the associated production of gauginos and gluinos at hadron colliders.
If supersymmetry exists at the electroweak scale, the cross section for
this process is expected to be large and observable at the Fermilab Tevatron
and/or the CERN LHC. It is enhanced by the large color charge of the gluino
and the (in many SUSY models) small mass of the light gauginos.
The cross section for $\neutralino_2$ production is the largest, because of its 
zino-like coupling.  The cross section for $\chargino_1$ is about equal to 
that of the $\neutralino_2$.  The NLO predictions are modestly larger than the 
LO values but considerably more stable theoretically.  

Work in the High Energy Physics Division at Argonne National Laboratory
is supported by the U.S. Department of Energy, Division of High Energy
Physics, under Contract W-31-109-ENG-38.  The authors are grateful for
correspondence with W.\ Beenakker and conversations with S.\ Mrenna.
T.\ Tait has benefitted from discussions with C.--P.\ Yuan.



\end{document}